\documentclass[aps,
prb,
preprintnumbers,
superscriptaddress,
footinbib,
amsfonts,
amssymb,
amsmath,
intlimits,
onecolumn,
]{revtex4-2}
\usepackage[utf8]{inputenc}
\usepackage[T2A]{fontenc}
\usepackage{float}
\usepackage{bm,latexsym,mathrsfs,enumerate,amsmath,amssymb}
\usepackage[table,x11names]{xcolor}
\usepackage[breaklinks=true,
unicode=true,
urlcolor = blue,
colorlinks = true,
citecolor = blue,
linkcolor = blue
]{hyperref}
\usepackage{graphicx}
\graphicspath{{./figs/}}

\usepackage{savesym}
\usepackage[most]{tcolorbox}
\savesymbol{comment}

\usepackage{todonotes}
\usepackage{easyReview}

\renewcommand{\vec}[1]{\bm{#1}}

\begin{document}

\preprint{This is the initial, pre-peer-review version of a manuscript published in Nature Nanotechnology.}

\title{Modern triad for magnetic material science:\\ geometric shape, electronic bands and spin textures}

\author{Denys~Makarov}
\email{d.makarov@hzdr.de}
\affiliation{Helmholtz-Zentrum Dresden-Rossendorf e.V., Institute of Ion Beam Physics and Materials Research, 01328 Dresden, Germany}

\author{Oleksandr~Pylypovskyi}
\email{o.pylypovskyi@hzdr.de}
\affiliation{Helmholtz-Zentrum Dresden-Rossendorf e.V., Institute of Ion Beam Physics and Materials Research, 01328 Dresden, Germany}
\affiliation{Kyiv Academic University, 03142 Kyiv, Ukraine}

\author{Rui~Xu}
\email{r.xu@hzdr.de}
\affiliation{Helmholtz-Zentrum Dresden-Rossendorf e.V., Institute of Ion Beam Physics and Materials Research, 01328 Dresden, Germany}

\author{Carmine~Ortix}
\email{cortix@unisa.it }
\affiliation{Dipartimento di Fisica ``E. R. Caianiello'', Universit\`{a} di Salerno, IT-84084 Fisciano (SA), Italy}

\date{October 30, 2025}

\begin{abstract}
To enable novel concepts in nanotechnologies, it is insightful to design materials with qualitatively new performances originating from recent discoveries in fundamental material science and physics. Curvilinear magnetism  emerged as a tool to design chiral and anisotropic responses of materials at the nanoscale using the effects of the geometric shape and topology of the object. This concept is distinct yet complementary to  material screening, which is primarily based on the optimization of intrinsic microscopic properties to modify magnetic textures for specific applications. Contemporary curvilinear magnetism focuses on mean-field micromagnetics, while the explicit contribution of the lattice structure and electronic degrees of freedom remains beyond consideration. In this review, we will emphasize on the effects which appear when this coupling is accounted for and that have resulted in novel concepts for the design of  ``metageometric'' materials. Further development of this field will provide appealing technological prospects for fundamental understanding of curvilinear magnetic nanostructures and their applications in energy efficient and scalable nanoelectronics.
\end{abstract}

\maketitle

\section{Introduction}

Ordering phenomena associated with symmetry breaking phase transitions are cornerstones of  modern condensed matter physics. They determine local (ferroic) orderings and the electromagnetic responses of materials. Until recently, research in solid state physics silently relied on the crystalline symmetry of a bulk material in \emph{planar} systems. This situation drastically changed over the last decade when imposing three-dimensionality (3D) in a low-dimensional system emerged as the only way to push further the performance of transistor architectures. This route has been applied in every technology including superconducting electronics, neuromorphic computing and energy efficient magnetoelectronics based on multiferroic materials and racetracks. Such a paradigm shift in technology highlighted the lack of fundamental understanding, in particular concerning the properties and responses of local order parameters in geometrically curved space. While the investigation of the effect of space curvature on quantum particles has been iniatiated more than 60~years ago~\cite{DeWitt57,Jensen71,daCosta81,Ortix15}, research on curvature effects in systems with long-range order as magnetism and superconductivity is still in its infancy~\cite{Gentile22,Makarov22a,Fomin22}.

\begin{figure}
    \centering
    \includegraphics[width=\linewidth]{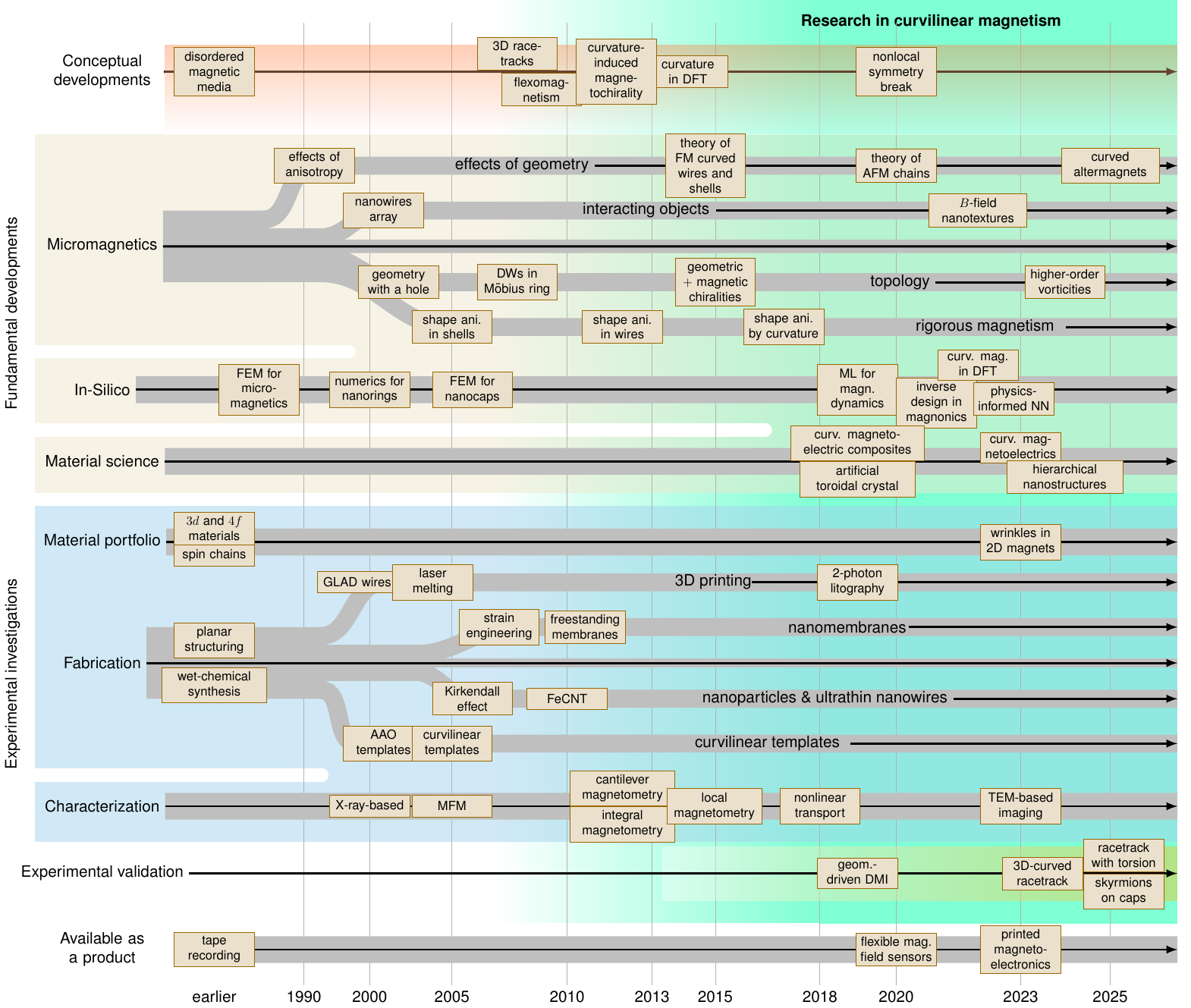}
    \caption{\textbf{Research timeline of curvilinear magnetism.} Numerous works on micromagnetics~\cite{VillainGuillot94,Hertel01,Saxena02,Kohn05a,Yoneya08,Slastikov12,Gaididei14,Pylypovskyi15b,Tretiakov17,Pylypovskyi20,Donnelly22,Volkov24a,Yershov25}, in-silico research~\cite{Fredkin87,LopezDiaz00,Ulbrich06,Kovacs19,Wang21d,Kovacs22,Kovacs22a,Edstrom22}, material science development~\cite{Volkov19b,Lehmann19,Ortix23,Bezsmertna24} and experimental techniques~\cite{Malac99,Nielsch01,Kruth04,Albrecht05,Gao06,Mei08,Tillier10,Wolny10a,Weber12,Streubel12b,Vock14,Maurenbrecher18,May19,Gu22,Feng22,ONeill23,Volkov23a,Jin25} have been supplemented by the conceptual developments in solid state~\cite{Andreev78,Parkin08,Eliseev09,Lukashev10,Stengel13,Hertel13a,Sheka20a} and entering market. Up to now, several theoretical predictions have been validated experimentally~\cite{Volkov19c,Fedorov24,Farinha25,Dugato25}. Abbreviations used: DFT (density functional theory), DWs (domain walls), FEM (finite-element mesh), ML (machine learning), GLAD (glancing-angle deposition), AAO (anode aluminum oxide), MFM (magnetic force microscopy), TEM (transition electron microscopy). }
    \label{fig:timeline}
\end{figure}

The major interest comes from the possibility to create and characterize at the nanoscale limit geometrically curved systems applying appropriate fabrication methods, which thus allow to design curvature effects in solid state objects. Such fabrication and characterization technologies, including 3D nanowriting~\cite{Fernandez-Pacheco20,Ladak22,Hoflich23,Fullerton24,Farinha25}, advanced x-ray and electron tomography/holography methods~\cite{Christensen24,Gubbiotti24}, matured over the last decade (Fig.~\ref{fig:timeline}). Considering the coherency requirement to the light sources for high-resolution 3D imaging, the next major boost in this research and technology field is expected with the next generation of synchrotron facilities. 

The 3D shape of nanoscale objects does not only change the structural geometry of planar systems, but also act as a source of point group symmetry breaking, including rotations and inversions,  which enables the design of material systems with a symmetry content low enough to allow new physical effects relevant for specific applications. In this respect, curvilinear solid state physics gets inspiration from the occurrence of effects, like the Rashba/Edelstein effect, which are limited to specific materials with  lowered crystalline symmetry content. This enables a rational design of geometries (i.e., curvatures and topology) of low dimensional architectures, 
providing discrete symmetry breakings at the structural level, analogue to the breaking of symmetries occurring at the lattice level~\cite{Ho21}. A further advantage of this concept is that it can even result in the appearance of much stronger classical analogs of quantum effects~\cite{Makushko24}.

Until recently, 
research on the effect of geometry in magnetic systems has been mainly limited to the simplest case of the coupling between the specific order parameter (e.g., magnetization) and geometric curvature (Fig.~\ref{fig:timeline}), while the coupling with the lattice degrees of freedom and the electronic band structure remained out of consideration. The latter results in the fact that many physical effects, including magnetoelectric coupling phenomena in insulators  were overlooked. The  coupling between geometry of the physical space with the electronic band structure for instance via mechanical strain and/or strain gradients, is expected to lead to novel material properties of curved systems in comparison with the planar counterparts. In thin films of topological insulators the coupling of inhomogeneous strain to the surface Dirac fermions can result in phase transition between topologically distinct insulating phases~\cite{Battilomo19a}. Sample geometry with nanoscale curvature also naturally influences the surface and boundary states in topological materials~\cite{Wang14d,Siu17}. Furthermore, the appearance of crystallographic defects, which are naturally expected in strained curvilinear geometries due to the fabrication process can be a source of stabilization of specific textures itself~\cite{Azhar22,Pylypovskyi23i}. In such a way, design of geometry provides versatile ways to tailor specific properties of solid state samples over their intrinsic material parameters, which, otherwise, may not support such states at all  (Fig.~\ref{fig:timeline}). These effects became relevant with the recent discovery of magnetic 2D materials with transition temperatures above room temperature~\cite{Huang17,Gong17,Burch18,Gibertini19,GrubivsicvCabo25}
which are very susceptible to strain effects~\cite{Edstrom22,Bagani24,Jin25}. Due to their low flexural rigidity and conformability, these materials are envisioned to come to the forefront of the curvilinear magnetism research. In addition to 2D magnets, effects of strain are relevant for antiferromagnet- and altermagnet-based spintronic devices for instance to set a pre-defined magnetic state~\cite{Feng20,Ni21,Chakraborty24,Li25a,Khodas25}. Furthermore, lowering the symmetry at the structure level via appropriate geometric deformations enables spin-texture-driven phenomena such as magnetoelectric responses due to geometrically-driven higher-order magnetic  multipoles~\cite{Ortix23,Pylypovskyi25}  (Fig.~\ref{fig:timeline}). This paves the way to a new class of magnetoelectric multiferroic materials which is referred to as curvilinear multiferroics where magnetoelectricity is enabled via a trilinear coupling term of toroidization or magnetoelectric monopolization,  magnetization, and electric polarization. 

The magnetostatics-driven phenomena are linked to the mapping effects of the geometric topology on the ferroic order parameter. At the local scale, they can govern the set of geometrically-induced energetically preferable states characterized by multiple chiral parameters~\cite{Volkov23a} or even prescribe the grouping of topologically non-trivial textures~\cite{Volkov24a}. On the mesoscale, magnetostatics provies several important features. Being non-local, it directly links different parts of the non-planar geometries and even neighboring mesoscopic structures~\cite{Hertel01,OBrien11,Brajuskovic21,Donnelly22}. As source of shape anisotropy, magnetostatics could induce anisotropic properties that follow (repeating) geometric patterns while the instrisic anisotropy is suppressed or negligible on the selected spatial scale. All together, the magnetostatically coupled, spatially repeated structural units of the selected geometry and intrinsic magnetic properties tuned by strain and coupled to the electronic degrees of freedom constitute a possibility of a \emph{metageometric material design}, where the anisotropic, chiral and multiferroic responses are given by the access to the specific spatial scale.

\section{Geometry, topology, strain and magnetic textures} 
\label{sec:geometry}

\begin{figure}
    \centering
    \includegraphics[width=\linewidth]{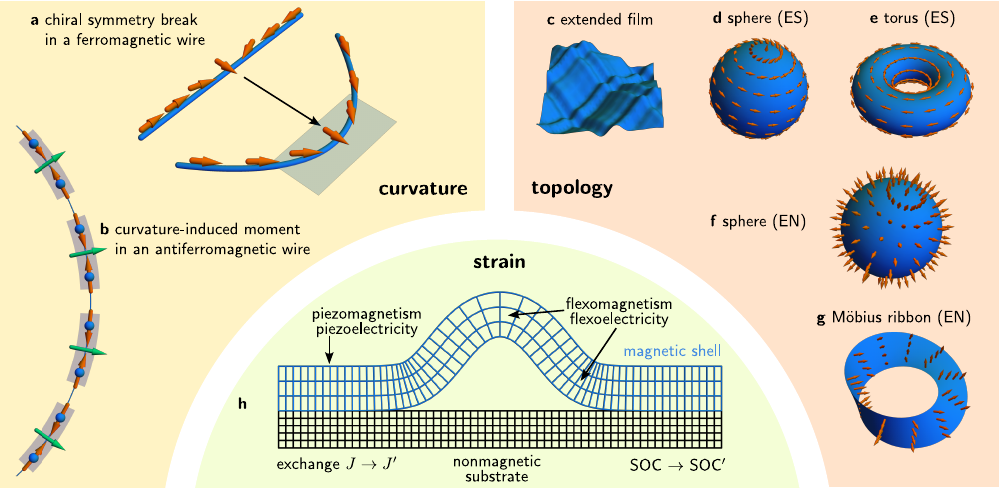}
    \caption{\textbf{Components of construction of a curvilinear magnet: (a,b)~geometric curvatures, (c--g)~topology and (h)~strain.} (a)~In a strain ferromagnetic wire chirality (magnetization direction shown by orange arrows) of a head-to-head domain wall is arbitrary. In a bent sample, the geometry-driven anisotropy fixes magnetization in the plane of bend (semitransparent blue plane), and gDMI fixes the outward chirality. (b)~Schematics of the geometry-induced magnetic uncompensation per dimer (green arrows, not in scale) in an antiferromagnetic spin chain when magnetic moments (orange arrows) are tangential to the chain. (c)~In comparison with an extended magnetically soft film, (d)~topology of a sphere imposes the presence at least of two vortices. Here, ``ES'' stands for the easy-surface magnetic anisotropy. (e)~Magnetically soft torus with a single hole is again topologically similar to the extended film~(c). (f)~Sphere's topology makes the radially-magnetized state topologically kin to a skyrmion in a plane sample. Here, ``EN'' stands for the easy-normal anisotropy with the axis along the surface normal~$\vec{n}$. (g)~Interlink of a M\"{o}bius ribbon with the EN anisotropy leads to the appearance of a topologically protected domain wall whose chirality is determined by the ribbon's chirality. (h)~Presence of strain in a magnetic shell with a wrinkle (blue) epitaxially grown on a substrate (black) results in change of exchange and spin-orbit (SOC) coupling effects, as well as induces piezo- and flexomagnetic effects in regions of a finite strain and strain gradients, respectively.}
    \label{fig:triple}
\end{figure}

Reasons of coupling between the physical shape of a sample and its magnetic order parameter include geometry-tracking phenomena, such as shape anisotropy, crystalline or surface anisotropy, or certain spin torques~\cite{Sheka22,Makarov22}. They offer novel means for a local modification of magnetic responses in curved geometries beyond boundary effects (e.g. vortex formation in thin nanodots), which have been established by the magnetism community for decades. In particular, in a thin bent shell, finite curvature breaks the inversion symmetry being mapped to Lifshitz invariants in the energy functional, and anisotropic terms arising due to inequivalence of local directions in a thin bent shell [see Box~\#1 and Fig.~\ref{fig:triple}(a)]. The main drivers of such local modifications of the magnetic properties are the exchange and Dzyaloshinskii--Moriya interaction of the material sample without explicit modification of the spin-orbit coupling (SOC) effects. The nonlocal chiral symmetry breaking appears due to the interaction of surface and volume magnetostatic charges that account for not only the shape of the object but also tracks the inequivalence between the top and bottom surfaces of the sample~\cite{Sheka20a}. The nonlocal magnetostatics induces chiral symmetry breaking in a curved sample and links the topology of the object geometry with the topology of the magnetization field~\cite{Volkov24a, Donnelly22,Bezsmertna24}. With the help of nonlocal magnetostatic interaction and/or currents, the geometric effects can be strongly pronounced in experiments demonstrating domain wall pinning at curvature maxima~\cite{Volkov19c} and motion of magnetic solitons along chiral racetracks \cite{Farinha25}. Furthermore, the geometry-induced symmetry breaking allows lifting the degeneracy of certain magnetic states and modify them with additional chiral twists.

\begin{tcolorbox}[title=Box~\#1. Differential geometry of thin magnetic shells, colframe=CadetBlue4,breakable]
\textbf{Geometry of thin shells.} Geometry of a thin shell can be understood via properties of a 2D manifold $\vec{\varsigma}(u,v)$ (``central surface'') with a coordinate system $\{u,v\}$ on $\vec{\varsigma}$. Two vectors tangential to $\vec{\varsigma}$ can be introduced as $\vec{g}_i = \partial_i \vec{\varsigma}$, $i=u,v$. Then, the unit normal to the central surface is $\vec{n}(u,v) = \vec{g}_u \times \vec{g}_v/|\vec{g}_u \times \vec{g}_v|$. The volume of the shell is
\begin{equation}\label{eq:shell-volume}
    \vec{r} = \vec{\varsigma}(u,v) + \zeta \vec{n}(u,v),\quad \zeta \in [-h/2,h/2],
\end{equation}
where $h$ is the shell thickness. The metric properties of $\vec{\varsigma}$ can be determined from two matrices, the first fundamental form $\|g_{ij}\| = \vec{g}_i \cdot \vec{g}_j$, and the second fundamental form $\|b_{ij}\| = \vec{n} \cdot \partial_j \vec{g}_i$. While $\|g_{ij}\|$ serves to calculate distances, $\|b_{ij}\|$ provides so-called principal directions $\vec{e}_{u,v}$ that are orthogonal vectors, and principal curvatures $\kappa_{u,v}$ as the eigenvectors and eigenvalues of the so-called shape operator of the second fundamental form, $\|h_{ij}\| = b_{ij}/\sqrt{g_{ii}g_{jj}}$. Description of the magnetic energy $E$ of the shell in the reference frame given by the principal directions provides a possibility restore the translational invariance of the geometry-tracking interactions and map the bends and twists of geometry into effective energy terms in $E$. In particular, the principal directions provide the axes for the geometry-driven anisotropy, while principal curvatures scale the geometry-driven anisotropy and Dzyaloshinskii--Moriya interaction (gDMI). A general description of local effects in curvilinear thin shell with the vector order parameter $\vec{m}(\vec{r})$ can be given via modified covariant derivatives applied to the components of $\vec{m}$ in the local reference frame, $\{m_u, m_v, m_n\}$~\cite{Sheka20a}. 

\textbf{Strain gradients.} Curvature of a thin magnetic shell also immediately implies the existence of strain gradients. Consider for instance the simple example of a magnetic shell which is translationally invariant in one-direction. The unit normal to the central surface is then only function of one coordinate, say $u$, and there is an associated single non-vanishing principal curvature $\kappa_u$. The components of the strain tensor $\|\epsilon_{ij}\|$ are easily computed as \begin{eqnarray}
\epsilon_{uu}&=&-\zeta \kappa_u(u) \\
\epsilon_{zz}&=&-\dfrac{\nu}{1-\nu} \epsilon_{uu}
\end{eqnarray}
where $\nu$ indicates the Poisson's ratio. It is important to note that such strain gradients are generated by bending in systems that preserve long-range crystalline order. Note also that in planar systems where strain gradients are generated near interfaces or surfaces can be mapped of as having a geometric curved shape according to the equations above. 

\textbf{Example.} On a surface of revolution like catenoid 
\begin{equation}
    \vec{\varsigma} = c  (\vec{\hat{x}} \cos v + \vec{\hat{y}} \sin v)\cosh \dfrac{u}{c} + \vec{\hat{z}}u,
\end{equation}
$u$ and $v$ can be chosen as an arclength of generatrix and an azimuthal angle, respectively. In this case, both fundamental forms are the diagonal matrices and $\|h_{ij}\| = \mathrm{diag}(-\kappa_0, \kappa_0)$ with $\kappa_0 = (1/c)\,\mathrm{sech}^2(u/c)$. Two coefficients of gDMI are $\mathcal{D}_{1,2} = \pm 2A\kappa_0$, where $A$ is the exchange stiffness. In particular case of $A = 10$\,pJ/m and curvature radius $1/\kappa_0 = 25$\,nm, gDMI is about 0.8\,mJ/m$^2$, which is comparable to the intrinsic DMI in Co/Pt multilayers. We note that in Eq.~\eqref{eq:shell-volume}, the thickness could be coordinate dependent being reflected in the energy functional as the effective geometric field determined by the gradient of sample's cross-section~\cite{Yershov23}. Non-local effects stemming from magnetostatics in ferromagnets are captured by the part of the magnetostatic volume charges (tangential charges)~\cite{Sheka20a} and non-local magnetostatic kernel. As the result, they can couple several magnetochiral parameters whitin one magnetic texture~\cite{Sheka20a,Volkov23a,Bezsmertna24}. 

\begin{center}
    \includegraphics[width=0.3\linewidth]{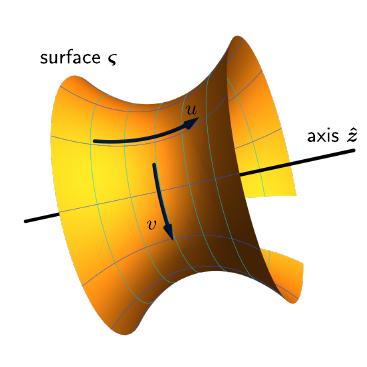}
\end{center}
\end{tcolorbox}

A particular feature of antiferromagnetic materials is a strong dependence on the structure of crystal lattice and spatial ordering of magnetic ions. In bulk materials this appears in a variety of phenomena relying on the staggered or noncollinear magnetic order such as magnetoelectricity and defect-driven topological textures. Curvilinear antiferromagnetic nanoarchitectures enable additional responses even at zero order of lattice deformation neglecting any strain effects. Unlike ferromagnets, the shape anisotropy in antiferromagnetic spin chains is of the hard-axis type, which allows the geometry-driven easy-axis anisotropy to be the decisive in setting the direction of spins~\cite{Pylypovskyi20}. It also removes the limiting curvature for the geometry-driven helimagnetic phase transition and allows tuning the spin-flop field by curvature. A local symmetry breaking induced by placement of spins along the curved path provides the weak ferromagnetic response even for the locally uniform textures [Fig.~\ref{fig:triple}(b)] and enables the homogeneous Dzyaloshinskii-Moriya response for the non-collinear textures~\cite{Pylypovskyi21e,Borysenko22, Salamone24}. Spin frustration effects in curvilinear antiferromagnets can be enhanced or suppressed by tailoring the curvature~\cite{Castillo-Sepulveda17, Pylypovskyi25}. The spatial distribution of curvature and defects enhances pinning of non-collinear antiferromagnetic textures making the energy landscape more diverse~\cite{Yershov22,Yershov22a,Jani24}. Beyond the pure action on the magnetic order parameter, the geometry alters transport properties enabling additional possibilities to probe and manipulate antiferromagnetic states~\cite{DasGupta22,Salamone24}.

Historically, the research on curvilinear magnetism typically considered local properties of the geometry, namely principal curvatures for surfaces or curvature and torsion for wires to modify responses of magnetic materials~[Fig.~\ref{fig:triple}(c)]. Understanding of these curvature-induced effects for ferro- and antiferromagnetically-coupled materials is by now rather complete. Recently, topology of the physical sample’s geometry was introduced as a key aspect to constrain the magnetization vector field [Fig.~\ref{fig:triple}(d--g)]. The conceptual strength of this approach is that the topological properties of a surface determine uniquely the number and type of magnetic solitons, which determines the relevance of the designed curvilinear architecture for 3D magnonic computing, reservoir computing, superconducting electronics, shaping magnetic near fields for biomedical applications~\cite{Medina-Sanchez17,Wang21b}. For instance, the surface of a magnetic $N$-pod without cross-linking is topologically equivalent to a sphere and hence can support $N$ vortices and $N-2$ antivortices (i.e., $2N-2$ magnetic solitons per object) as a ground state of a magnetically soft ferromagnet [Fig.~\ref{fig:triple}(d)]~\cite{Sloika17,Volkov24a}. Even more interesting that relying on 3D nanoscale direct writing~\cite{Fernandez-Pacheco20,Hoflich23,Fullerton24}, it is possible to realise objects with the topology of $N$-torus (i.e., torus with $N$ holes) [Fig.~\ref{fig:triple}(e)]. For these necessarily 3D-shaped wireframe geometries, the prevailing type of magnetic solitons is antivortices rather than vortices (which are very usual in 2D geometries)~\cite{Carvalho-Santos10,Vojkovic17}. For instance, 4-torus supports 6 antivortices only. At the same time, buckyball-shaped wireframes can support an excess of 60 antivortices within one spatially confined 3D magnetic object~\cite{Donnelly15,Cheenikundil21,Cheenikundil22,Pip22}. At the same time, the bulk magnetic textures in toroidal geometries may correspond to hopfions~\cite{Castillo-Sepulveda21}.

For the magnetically hard nanomagnets whose magnetic easy axis tracks surface normal [Fig.~\ref{fig:triple}(f,g)], the Gauss-Bonnet theorem applied to the Gauss map (normal vector field) reveals the relation between the topological charge of the magnetic texture and topology of the sample. The Gauss map provides a ``shift'' for the topological number of magnetic texture which is zero for the extended films and 1-tori, but finite for spheres and $N$-tori. As the consequence, the true topologically trivial state can host topological defects locally at the curved surface with an example of a single skyrmion on a magnetic sphere~\cite{Kravchuk16}.

Transforming a planar system into a 3D geometry often leads to the modification of local strains and imposes strain gradients [Fig.~\ref{fig:triple}(h)]. This leads to complex responses beyond the chiral and anisotropic ones, such as giant flexomagnetic~\cite{Qiao24} and flexomagnetoelectric~\cite{Shen18b} effects, or, even, time-reversal symmetry breaking enabling magnetic responses~\cite{Tao14}. In 2D magnets, the effects of strain and strain gradient due to curvilinear geometry become of major relevance as was demonstrated by~\citet{Edstrom22} for the case of nanotubes of CrI$_3$ and studied experimentally for different families of curved 2D magnets~\cite{Thiel19,Bagani24,Jin25a,Jin25}. Admittedly, although simulation packages that can treat effects of mechanical strain in curved magnetic objects do exist~\cite{Challab21,Chiroli23}, the effects of strain and strain gradient are not taken into account in the modern theory of curvilinear magnetism. Therefore, it can be expected that there will be additional curvature-induced effects which will be unveiled as soon as a more complete theory of (inhomogeneously) strained curvilinear architectures will be developed. The pioneering work by~\citet{Edstrom22} already pointed out that the coupling between strain gradients and local order parameter is strongly dependent on the presence of SOC which has a major implication on the final energetically preferred magnetic state. This means that prediction of magnetic states in highly curved nanoarchitectures would necessarily require multiscale analysis, where phenomenology of the curvilinear magnetism should be complemented with the results of \textit{ab initio} calculations providing curvature/strain-modified coupling coefficients.

The effects of strain gradient are tensorial. Interestingly, the presence of inhomogenous strain always imply geometric curvature and, in fact, the effects of strain gradient and geometric curvature often cooperate with each other~\cite{Ortix11}, see Box~\#1. In this respect, the collected know-how in the curvilinear magnetism community can be efficiently used to qualitatively discuss the effects of inhomogeneous strain in planar but also geometrically curved magnetic nanoarchitectures.

\section{Curvilinear magnetoelectrics}
\label{sec:magnetoelectrics}

\begin{figure}
    \centering
   \includegraphics[width=\linewidth]{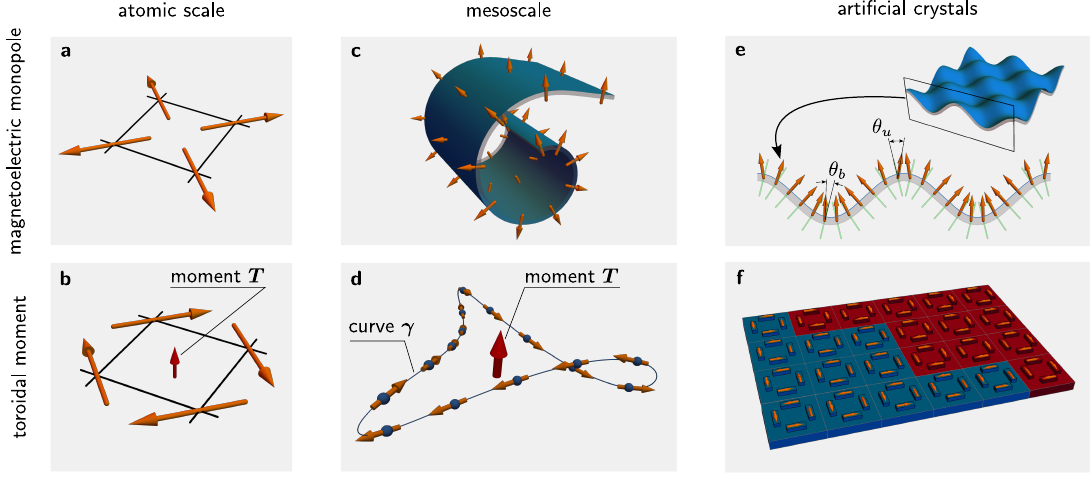}
    \caption{\textbf{Magnetoelectric monopoles (a,c,e) and toroidal moments (b,d,f) at different length scales.} Atomic scale ordering of the (a)~magnetoelectric monopole $A$ (pseudoscalar) and (b)~toroidal moment $\vec{T}$ at the square lattice formed by magnetic moments (orange arrows). (c)~Rolled-up thin magnetoelectric film with the out-of-plane anisotropy has a finite monopolar moment in addition to non-extensive toroidal moment (here and in the following panels, magnetization direction is shown by orange lines). (d)~A ferromagnetic spin chain arranged along a closed space curve $\vec{\gamma}$ with the easy magnetic axis along tangential direction of $\vec{\gamma}$ forms a finite toroidal moment $\vec{T}$ perpendicular to the chain's plane~\cite{Pylypovskyi25}. (e)~Curvilinear magnetoelectric membrane with a finite monopolization due to broken time and space inversion caused by the intrinsic DMI: angles between magnetization (orange arrows) and normal (green lines) are different at the upper and bottom parts, $\theta_b \neq \theta_u$~\cite{Ortix23}. (f)~Artificial toroidal crystal formed by the uniformly magnetized mesostructures at the top of plane magnetically soft film~\cite{Lehmann19}. Two different domains corresponding to the opposite directions of $\vec{T}$ are shown by blue and red colors.}
    \label{fig:magnetoelectrics}
\end{figure}

In contrast to extensive explorations of the effect of curvature, shape topology and strain on magnetic states, active investigations of such effects for magnetoelectric functionalities is very new and goes back to 2023~\cite{Ortix23,Pylypovskyi25}. The concept behind curvilinear magnetoelectric has emerged primarily in geometrically curved antiferromagnets but can be geneneralized also to ferromagnetic and helimagnetic curvilinear architectures. Generally speaking, the class of magnetoelectrics comprises magnetically ordered states characterized by higher multipole orders, such as magnetic quadrupole or magnetic toroidal dipole order, which are each associated with distinct patterns of the magnetoelectric response~\cite{Spaldin13,Thole16,Gao18}. These higher order multipoles generally appear when the magnetically ordered state breaks time-reversal and spatial inversion symmetry, but preserves their product, the so-called anti-inversion. This forbids ferroelectricity but allows for a linear magnetization response to an applied electric field. Materials exhibiting such multipolar order and the associated magnetoelectric responses have seen a surge of interest, and have come under intense scrutiny in recent years. From a fundamental point of view, much work has focused on the proper categorization of distinct types of magnets and their magnetoelectric responses, as well as in defining the magnetoelectric multipoles as bulk quantity expressed in terms of quantum mechanical wavefunctions~\cite{Bhowal22}. From the material point of view, the focus has been in finding specific realizations of higher-order magnetoelectric multipoles. It has been shown that certain bulk crystals, as for instance the lithium transition metal phosphates, can host magnetoelectric monopolization which immediately equips them with a magnetoelectric capability. Other transition metal compounds have been instead found to host a finite toroidization. In these materials, the magnetic sites realize non-collinear spin textures {[see Fig.~\ref{fig:magnetoelectrics}(a,b)]}, thus necessitating for the presence of strong spin-orbit coupling. Curved magnetic architectures offer an alternative solution since the shape itself can create a non-collinear texture with a net zero moment, a priori even in the complete absence of spin-orbit coupling. Consider the case of a cylindrical nanotube: in the case of an easy-axis ferromagnet, the ensuing radial magnetization realizes a prime example of a magnetoelectric monopole moment [Fig.~\ref{fig:magnetoelectrics}(c)]. Similarly for an easy-surface ferromagnet a magnetic toroidal moment is created. More complex geometries where antiinversion is naturally broken might harbor the concomitant presence of more magnetoelectric multipoles {[Fig.~\ref{fig:magnetoelectrics}c]}. This is realized in the case of a spirally-wrapped nanotube which possesses a subleading magnetic toroidal moment with radial magnetization. The presence of a toroidal moment together with a magnetoelectric monopole is clearly expected to lead to richer patterns of induced electric polarization, and hence augmented magnetoelectric capabilities. The interplay between curvature effects and the topology of the magnetic architectures is also relevant for magnetoelectric capabilities. In closed-loop epicycles {[Fig.~\ref{fig:magnetoelectrics}d]}, it has been shown for instance that the spin texture spontaneously selects a purely toroidal configuration with the toroidal moment pinned to the rotation axis of the epicycle. 

It is important to point out that the magnetoelectric multipoles realized in such isolated architectures does not possess the ferroic nature of an additional order parameter, which in a Landau energy functional [see Box~\#2] couples to magnetization and electric polarization in trilinear terms. Magnetoelectric multipole ordered ground states with domains that can be oriented by conjugate fields can be found in superlattices – artificial crystals – realizing at the mesoscopic scale the atomic non-collinear magnetic textures of strongly correlated compounds. Such artificial crystals {[Fig.~\ref{fig:magnetoelectrics}f]} have been first obtained using permalloy planar nanomagnets~\cite{Lehmann19} with ferrotoroidic ground state. Importantly, it is possible also to achieve all-geometric superlattices with magnetoelectric multipole order. In two-dimensional magnets with, e.g. thermally or strain-activated, periodic ripples it has been shown~\cite{Ortix23} that curvature-induced effects on the pristine ferromagnetic arrangement of the material {[Fig.~\ref{fig:magnetoelectrics}e]} together with the intrinsic DMI coupling lead to a reorganization of the magnetic texture that breaks inversion symmetry with the compensated part of the magnetic texture equipped with a finite magnetoelectric monopole. The net monopolization in Janus Cr(I,Br)$_3$ has been estimated to be comparable to the monopolization predicted in the bulk lithium manganese phosphate. Furthermore, the presence of a net uncompensated magnetization in each rippled period implies that such superlattices of 2D magnets develop a ferroelectric polarization at the onset of magnetic order.  {In such a way, geometric superlattices enable magnetoelectric responses on the mesoscale complementing the the purely magnetic properties at the nanoscale. The hierarchy of different spatial scales offered by 3D curvilinear superlattices provides an appealing possibility to tailor magnetoelectric ordering of the required symmetry.}  Therefore, together with magnetoelectric functionalities, these artificial crystals can be categorized as proper multiferroics. The ensuing control of nonvolatile magnetization relying on an applied electric field paves the way towards prospective energy efficient magnetic memory and logic devices.

\begin{tcolorbox}[title=Box~\#2. Landau theory of magnetoelectric coupling in systems with higher multipole order, colframe=CadetBlue4,breakable]
{To show the magnetoelectric functionalities of magnetic systems characterized by higher-order multipole moment, we can} 
consider a Landau free energy expansion 
{involving the electric polarization, the magnetization and the higher-order magnetoelectric multipole, which, we choose here to correspond to the magnetoelectric monopolization $\mathcal{M}$.}
Additionally, the free energy expansion accounts for the coupling of electrical polarization and magnetization to the electric and magnetic field respectively. Finally, {and due to the transformation properties of the order parameters under inversion and time-reversal symmetry} there is 
a trilinear coupling among electric polarization, magnetization and monopolization.
The free energy reads as
\begin{equation} 
{\mathcal F}=  \dfrac{1}{2 \epsilon} {\vec P}^2 - {\vec P} \cdot {\vec E} +  \dfrac{1}{2 \chi} {\vec M}^2 - {\vec M} \cdot {\vec H} + \alpha \mathcal{M} {\vec P} \cdot {\vec M} + \dfrac{1}{2\chi_\mathcal{M}}\mathcal{M}^2 + \beta_\mathcal{M} \mathcal{M}^4 
\end{equation}
In the equation above the susceptibility $\chi_{\mathcal{M}}$ is considered temperature-dependent and diverges at the {magnetic ordering transition}
temperature. In addition, $\epsilon$ and $\chi$  are the electric and magnetic susceptibility while $c$ quantifies the strength of the magnetoelectric coupling. 
{Minimization of the free energy with respect to electric polarization and magentization gives rise to the two coupled equations}
\begin{equation}
 {\vec P}=\epsilon {\vec E} - \epsilon \alpha \mathcal{M} {\vec M}  
\end{equation}
\begin{equation}
{\vec M}=\chi {\vec H}- \chi \alpha \mathcal{M} {\vec P} 
\end{equation}
The magnetoelectric functionality is instead demonstrated by the fact that combining the two equations (and at leading order in $\mathcal{M}$) we have that an electric polarization is generated by an external planar magnetic field while a magnetization is generated by an external electric field according to 
\begin{eqnarray}
{\vec P}&=& \epsilon {\vec E} - \epsilon \chi \alpha \mathcal{M} {\vec H} \\ 
{\vec M}&=& \chi {\vec H} -\epsilon \chi \alpha \mathcal{M} {\vec P} 
\end{eqnarray}
A similar analysis can be performed for magnetic systems with higher order multipole order in the form of toroidization or quadrupolization. The only difference is in the form of trilinear coupling. 
\end{tcolorbox}

\section{Experimental realizations and observations}
\label{sec:experiment}

There are numerous experimental methods to fabricate high quality curvilinear architectures. Those include focused electron beam induced deposition~\cite{Fullerton24,Skoric20}, direct laser writing~\cite{Williams18,Sahoo21,Askey24,Farinha25}, template-assisted deposition~\cite{Bochmann18,Bezsmertna24}, controllable bending~\cite{Streubel14,Karnaushenko18} and wrinkling~\cite{Jin25}. In recent years, it became clear that technological advances will be possible if a scalable method for providing curvilinear structures will be proposed. In particular, strain engineering and template assisted deposition are positioned as such a technology. Through the selective etching of underlying sacrificial layers, the active films can be released and driven by built-in strain to spontaneously roll up into tubular architectures. These procedures are fully compatible with standard semiconductor processes, ensuring seamless integration into industrial-scale manufacturing. In addition, template-assisted fabrication approaches have emerged as a versatile and scalable route for constructing arrays of functional 3D structures owing to their broad adaptability to a variety of material deposition strategies~\cite{Xu22d,Bezsmertna24}. Nevertheless, achieving precise geometry control over the large-area 3D arrays remains beyond the reach of all the aforementioned methods. Such a constraint fundamentally limits the exploitation of shape- and topology-dependent phenomena, posing a significant obstacle to the realization of next-generation magnetic devices.

The possibility to realize free-standing nanomembranes of these large area materials, which can be transferred to any substrate of interest is another key pre-requisite for successful technological explorations especially for magnetization dynamics characterization. The possibility to transfer curvilinear magnetic nanomembranes to substrates, which are x-ray and electron beam transparent allowed 3D imaging magnetic states using holography and tomography approaches. To this end, already now magnetic states of curvilinear magnets can be successfully addressed based on electron holography and tomography as well as different synchrotron-based techniques where coherent imaging modalities provide a major boost offering high resolution tomography studies~\cite{Christensen24}. The availability of next generation of synchrotrons with improved coherence will definitely play an important role in the experimental validation of the theoretical concepts and promote realization of next-generation 3D magnetic nanodevices. Other experimental techniques should be explored by the community. For instance, it is important to consider methods which can probe second order processes like optical second harmonic generation yet applied to near field microscopies~\cite{Chauleau24} as well as transport-based methods like nonlinear Hall effect~\cite{Du21,Ortix21a,SuarezRodriguez25}. Those methods could become handy in addressing higher-order multipolar
states of curvilinear magnetoelectrics. In addition, high resolution microscopies are needed to visualize complex noncollinear magnetic states in these geometrically curved objects. Here we refer to magnetic force microscopy~\cite{Feng22,Dugato25} and potentially scanning nitrogen vacancy magnetometry~\cite{Rovny24} yet with improved resolution to 10\,nm and better. As these imaging tools continue to evolve in spatial and temporal resolution, they will play a critical role in guiding the design, fabrication, and theoretical modelling of curvilinear devices.

Application potential of geometrically curved magnetic architectures including thin films, nanocaps and nanowires currently being explored as spin-wave filters~\cite{Balhorn10,Otalora16,Korber21a,Brevis24}, bit-patterned recording media~\cite{Albrecht05,Albrecht15,Yang16,Tejo20}, percolated recording media~\cite{Rahman07,Laughlin07,Brombacher09,Grobis11}, high-speed racetrack memory devices~\cite{Parkin08,Yan10,Fedorov24,Farinha25}, magnetic logic~\cite{Allwood05,Luo20}, spatially configured magnetic field sensors~\cite{Streubel14a,Karnaushenko15c,Meng21,Ha21,Becker22}, magnetic soft robotics~\cite{Kim18a,Hu18,Zhang21b,Dreyfus24,Xu24a}, magnetic micro/nanorobotics~\cite{Ghosh09,Zhou21,Dong22}, on-skin interactive electronics relying on thin films~\cite{Wang16d,Loong16,Kondo20,Bezsmertna25b,Makushko25,Pan25} as well as printed magnetic composites~\cite{Xu25} with appealing self-healing performance~\cite{Bandodkar16,Wang19h,Cheng21,Xu22a}  and recyclability~\cite{Wang24c}. This opens perspectives for magnetoelectronics in smart wearables~\cite{Chen21a,Wan21,Zhao21b,Banerjee23,Gong25a,Lugoda25}, interactive printed electronics~\cite{Cox13,Gupta22,Xu25}, health care and medicals~\cite{Nam03,Karnaushenko15a,Krishna16a,Rizzi17,Xu20c,Xu20d}, energy harvesting~\cite{Zhou21a,Chen21b,Li22c,Xiao23}, and motivates further explorations towards the realization of eco-sustainable magnetic field sensing relying on biocompatible and biodegradable materials. 

\begin{figure}
    \centering
    \includegraphics[width=\linewidth]{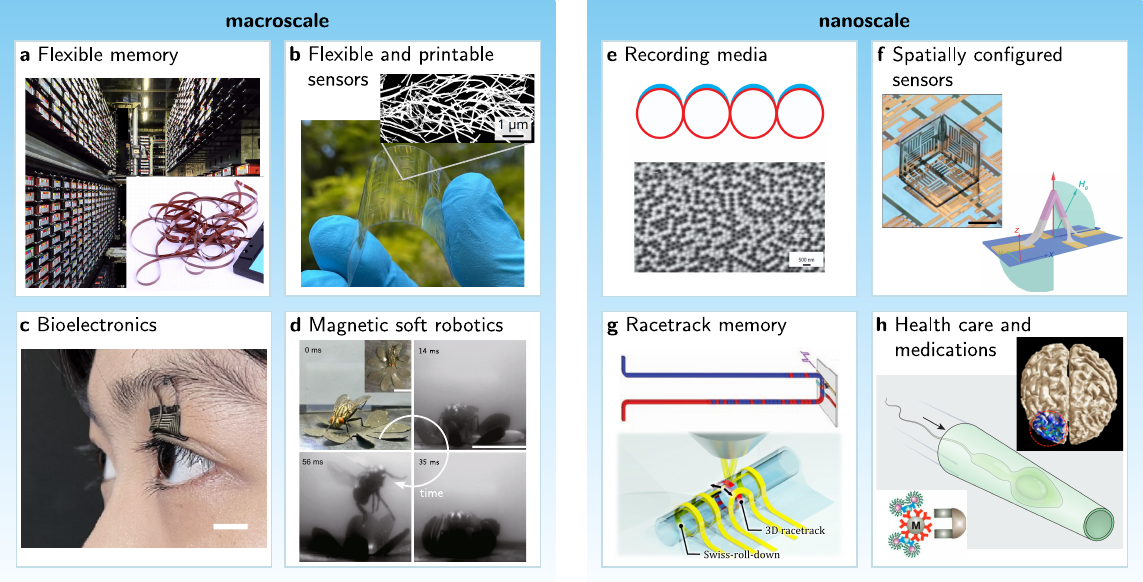}
    \caption{\textbf{Representative applications of curvilinear magnetic architectures in functional devices, spanning from macroscopic implementations (a--d) to nanoscale systems (e--h).} (a)~Commercial magnetic tapes realized through flexible magnetic substrates, illustrating large-scale utility of curved architectures. (b)~Printable magnetoresistive sensors based on aligned nanowires, featuring mechanical flexibility and optical transparency, positioning them as potential candidates for wearable electronics.  (c)~Mechanically adaptive biosensors for precise decoding of fatigue levels. (d)~Soft robotic actuators exhibiting ultrafast mechanical response through integration of magnetic composites.  (e)~High-density magnetic recording media utilizing multilayered magnetic films deposited on nanospheres.  (f)~Three-dimensionally reconfigurable magnetic field sensors capable of adaptive spatial response.   (g)~Conceptual schematics of racetrack memory based on ferromagnetic nanowires (top) and self-assembled tubular 3D architectures (bottom). (h)~Biomedical and healthcare applications, including targeted drug delivery, hyperthermia, magnetogenetics~\cite{Choi24} and magnetoelectric neurostimulation enabled by curvilinear magnetic structures.
    (a)~Back image: adapted under Creative Commons CC BY 2.0 license (https://creativecommons.org/licenses/by/2.0/deed.en), \textcopyright{} The author~\cite{Franganillo22}. Inset: Reproduced under Creative Commons CC BY 2.0 license (https://creativecommons.org/licenses/by/2.0/deed.en)~\cite{Reckmann15}. (b)~Adapted under Creative Commons CC BY 4.0 license (https://creativecommons.org/licenses/by/4.0/), \textcopyright{}~The~Authors~\cite{Xu25}. (c)~Reused with permission~\cite{Xu25a}. (d)~Adapted under a Creative Commons Attribution 4.0 International License (https://creativecommons.org/licenses/by/4.0/), \textcopyright{}~The~Authors~\cite{Wang20e}. (e)~Adapted with permission~\cite{Albrecht05}. (f)~Adapted under a Creative Commons Attribution 4.0 International License (https://creativecommons.org/licenses/by/4.0/). Left panel: \textcopyright{}~The~Authors~\cite{Becker22}. Right panel: \textcopyright{}~The~Authors~\cite{Meng21}. (g)~Top panel: adapted with permission~\cite{Parkin08}. Bottom panel: Adapted under a Creative Commons Attribution 4.0 International License (https://creativecommons.org/licenses/by/4.0/), \textcopyright{}~The~Authors~\cite{Fedorov24}. (h)~Central image: reproduced with permission~\cite{Medina-Sanchez17}. Left inset: reproduced with permission~\cite{Nam03}. Right inset: Adapted under a Creative Commons Attribution 4.0 International License (https://creativecommons.org/licenses/by/4.0/), \textcopyright{}~The~Authors~\cite{Guduru18}. 
}
    \label{fig:experiment}
\end{figure}

\section{Outlook}
\label{sec:outlook}

On the fundamental side, there are numerous effects to be explored, which originate from the curvature-induced breaking of spatial inversion symmetry and other point-group discrete symmetries. In particular, any curved flakes of 2D magnets necessarily develop inhomogeneous distributions of strain, and potentially, sizeable strain gradients that impact their magnetic state. In particular, curvature gives a possibility to adjust the symmetry of the primary order parameter by engineering the geometry of the object. Recently, in addition to the established activities on curvilinear ferro- and antiferromagnets, the exploration of effects of geometry in bent altermagnets was initiated~\cite{Yershov25}. To this end, curved altermagnetic shells share another possibility for tailoring uncompensated magnetization determined by the curvature. Furthermore, a recent theory has put forward that two-dimensional antiferromagnets in buckled lattices can display a topological magnetoelectric effect due to higher order multipoles~\cite{Venderbos25}. This type of magnetoelectric responses in solids have attracted a great deal of attention in the context of topological insulators~\cite{Hasan10,Qi11}. There are strong activities related to hybrids, e.g. superconductor/magnet heterostructures~\cite{Makarov22}, which address curvature-induced long-range supercurrents~\cite{Salamone21}, curvature-controlled proximity effects~\cite{Salamone22} among other appealing research directions. 
Benefiting from the specific topology of $N$-pod and $N$-torus structures,  strong gradients of the magnetic near field can be realized~\cite{Volkov24a}. This renders these objects relevant as components of smart micromachines for their navigation and localization in microsurgery, drug delivery, and artificial fertilization tasks. Furthermore, complex magnetic stray field patterns enable the trapping of magnetically-functionalized objects in biomedical screening assays. Undoubtedly, these topics will be in focus of the magnetism community in next years.

\begin{figure}
    \centering
    \includegraphics[width=\linewidth]{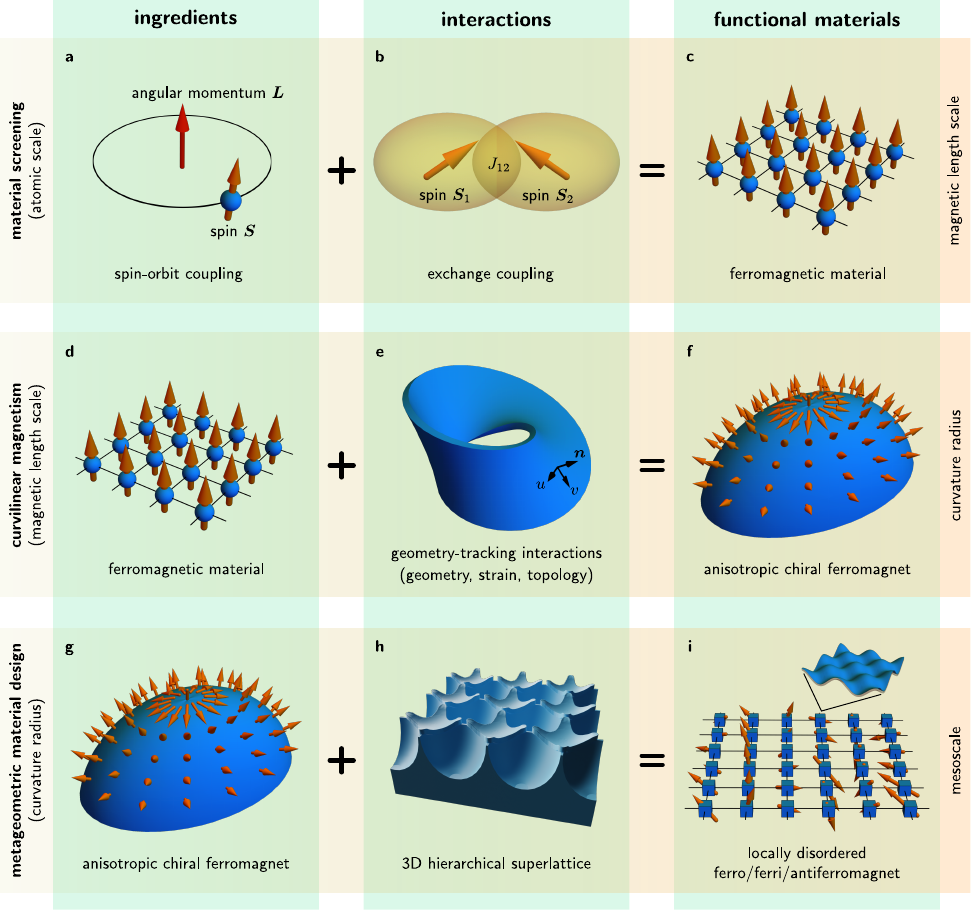}
    \caption{\textbf{Approaches in material science from the atomic to mesoscale.} A classical methodology relies on (a)~the design of the local atomic environment and (b)~coupling between spins to achieve (c)~the magnetic material with the desired properties. This material, e.g. specific ferromagnet serves as (d)~the ingredient for the modern field of curvilinear magnetism. (e)~Using toolbox that includes geometry-tracking interactions (shape anisotropy, strain and others) for the sample with the specific topology and geometry, (f)~the anisotropic chiral ferromagnet that could host, e.g., curvature-stabilized skyrmions, is obtained. Intrinsically, the material remains the same. Future of curvilinear magnetism: (g)~geometrically curved samples serve as the building blocks of (h)~3D superlattices with hierarchical structuring~\cite{Bezsmertna24} of their shapes and local responses. (i)~By spatial averaging of the superlattice responses over certain areas, they are rendered as locally disordered magnetic materials~\cite{Andreev78,Andreev80} on the mesoscale, where the macroscopic properties are determined by the geometric and magnetic symmetry of their building blocks~(g).
    }
    \label{fig:mesoscale}
\end{figure}

Beyond individual objects, we envision major developments enabled by the possibility to realize large area (wafer scale; 300\,mm wafers) defect free curvilinear templates offering scalable fabrication of metageometric structural units~\cite{Bezsmertna24} hosting numerous magnetic solitons in the ground state. The final symmetry properties of each structural unit are determined by the triad of specific geometry with its associated strain and strain gradients, electronic degrees of freedom 
and magnetic texture which is either pinned, or represents the ground state of the unit. It is possible to show that on the mesoscale the properties of these extended curvilinear templates hosting magnetic solitons within each unit cells yet at random positions. {Properties of such mesoscale structures} could be mapped on planar disordered magnets~\cite{Andreev78,Andreev80}. This possibility offers an appealing novel route towards the design of functional materials where the smallest constituent of the material will be represented by a curvilinear unit cell of the template, see Fig.~\ref{fig:mesoscale}. This is a new twist in the evolution of curvilinear magnetism, which for the last decade was positioned as a new approach to the material science enabling the design of chiral and anisotropic responses starting from say ferro- or antiferromagnetic materials. The geometry tracking interactions as well as the sample topology enabled the possibility to modify properties of conventional achiral magnets to act as anisotropic chiral materials. The concept of curvilinear magnetism postulated a major shift in the design of functional materials which was historically based on the targeted tailoring of material properties relying on material screening, alloying, doping etc. 

Although geometry gave us the possibility to design magnetic responses at will, curvatures of individual objects did not allow us to alter the effective exchange coupling of the material. The planar counterpart of a curvilinear ferromagnetic material was necessarily ferromagnetic material yet with locally tailored chiral and anisotropic responses. This was the major shortcoming of the curvilinear magnetism compared to the classical material screening approach which has the possibility to change the parent ferromagnetic coupling by, for einstance, doping into an antiferromagnetic coupling.
The hierarchical curvilinear {magnetic superlattices} equip us with this possibility and, in addition, restore the translational invariance of the lattice. Such templates offer the same functionality in the design of functional magnetic materials yet where the building blocks are artificially designed by geometry.
We envision that the prospective major discoveries in curvilinear magnetism will be done with hierarchical templates.

Fig.~\ref{fig:mesoscale} {summarizes} different approaches in material science. Classically, {material screening} (Fig.~\ref{fig:mesoscale}a—c) operates with the exchange-coupled atoms embedded into the given crystallographic environment. Commonly, the translational invariance of the crystal lattice within the sample is assumed. Introduction of the geometric curvature (Fig.~\ref{fig:mesoscale}d,e) with the geometry-tracking interaction breaks the local translational invariance, but provides an opportunity to modify the local anisotropic and chiral properties of the sample (Fig.~\ref{fig:mesoscale}f). As a next leap in the development of curvilinear magnetism we envision the use of curvilinear magnets as building blocks of novel functional materials (Fig.~\ref{fig:mesoscale}g--i). Together with the hierarchical structuring and coupling (either direct interface, or by stray fields, Fig.~\ref{fig:mesoscale}h), a ``new'' material on the mesoscale can be obtained (Fig.~\ref{fig:mesoscale}i). {The mesoscale responses can include geometry-driven magnetoelectricity as well as artificial anisotropic and chiral symmetries.} We refer to this approach as the metageometric material design. This vision strongly benefited from numerous foundational works on multipolar analysis of magnets~\cite{Andreev78,Andreev80}, fabrication and analysis of extended curvilinear superlattices~\cite{Korner09,Korner14,Tretiakov17,Lehmann19,Bezsmertna24}, complex 3D wireframes~\cite{Llandro20,Ladak22,Koshikawa23} and artificial spin ice~\cite{Heyderman21a}. 

In this respect, the concept of curvilinear magnetoelectric materials introduced in Section~\ref{sec:geometry} entirely relies on the availability of highly periodic curvilinear templates to realize toroidal ordering-enabled magnetoelectricity. As mentioned before, the topic of magnetoelectricity induced by geometry has been introduced only in 2023~\cite{Ortix23}. The theoretical description has been so far limited to the computation of the higher-order magnetic moments without any prediction on the strength of the ensuing magnetoelectric effect. It has been shown that in buckled lattices -- rippled membranes at the atomic scale -- the concomitant presence of the non-symmorphic twofold screw rotation symmetry and antiinversion symmetry guarantees the presence of symmetry protected massless Dirac fermions~\cite{Venderbos25}. Their inherent topology is reflected in a topological part of the spin magnetoelectric monopole and of the spin magnetoelectric polarizability. A promising future direction in this regard is to explore whether and how such topological magnetoelectric response can be observed in experimentally relevant larger scale superlattices which share the same non-symmorphic symmetry of buckled lattices. 

The current theory of curvilinear magnetoelectrics is purely phenomenological and based on symmetry analysis. In this respect, the design rules discussed in Section~\ref{sec:magnetoelectrics} related to the choice of material properties and sample geometries should be considered only as necessary but not sufficient conditions to observe magnetoelectric responses in the experimental realm. To assess the relevance of magnetoelectric functionality for experimental observations and real world applications, the next essential step would be to quantify the magnetoelectric polarizability with the additional knowledge of the electronic properties of the material system at hand. Gauge-invariant expressions for spin multipolization and the ensuing magnetoelectric polarizability using Bloch electronic wavefunctions have been theoretically derived using the semiclassical theory of electron dynamics~\cite{Gao18}, and successfully applied to magnetic skyrmions~\cite{Bhowal22}.

We note that experimental realizations of curvilinear magnetoelectrics are still pending. However, based on the available toolset of fabrication and characterization methods outlined in Section~\ref{sec:magnetoelectrics}, we are confident that it is a matter of the near future when the first experimental reports on curvilinear magnetoelectrics will be presented. The most exciting will be to explore the possibility to manipulate magnetic states of nanowires with toroidal ordering relying on an external electric field. This can enable a new paradigm in current concepts of magnetic racetrack memory based on 3D-shaped quasi-1D architectures~\cite{Farinha25} and magnetoelectric spin-orbit logic (MESO)~\cite{Manipatruni19}. 

Arguably, a concept can be considered impactful if in addition to appealing ideas for breakthrough fundamental research there are also prospects for technology transfer. To this end, current activities on curvilinear magnetism went beyond fundamental science. Indeed, there are ongoing technology transfer activities on flexible magnetoelectronics based on geometrically curved magnetic thin films and wires, magnetic composites for printed magnetic field sensors, methodology towards electronic conditioning of mechanically flexible and printed spintronic devices. These commercialization activities benefit from more than 30 years of research on flexible magnetoelectronics~\cite{Bermudez21}, which currently resulted in the possibility to fabricate magnetic field sensors in a fully automated batch processing over 300 mm wafers as well as prepare printed magnetic field sensors using scalable screen printing methods. In contrast to these macroscale devices, the technology transfer of curvilinear micrometer and nanometer scale devices remain limited. We hope that new concepts of curvilinear magnetoelectric materials, hierarchical architectures discussed in this review will stimulate further critical thinking of the community towards application-oriented explorations.

\section*{Acknowledgements}

The project is supported via the German Research Foundation (grants \#MA5144/33–1, \#MA5144/37–1), European Commission (project REGO; ID: 101070066), and ERC grant 3DmultiFerro (Project number: 101141331). C.O. acknowledges partial support by the Italian Ministry of Foreign Affairs and International Cooperation PGR12351 (ULTRAQMAT) and from PNRR MUR Project No. PE0000023-NQSTI (TOPQIN).

\end{document}